\documentclass[sigconf, screen=true, natbib=true]{acmart}

\acmSubmissionID{31}

\usepackage{xcolor,pifont}
\usepackage{pifont}
\usepackage{makecell}
\usepackage{graphicx}
\usepackage[skip=2pt]{caption}
\usepackage{subcaption}
\usepackage{hyperref}
\usepackage{acronym}
\usepackage{xspace}
\usepackage[inline]{enumitem}
\usepackage{multirow}
\usepackage{booktabs}
\usepackage[T1]{fontenc}
\usepackage{tikz}
\usepackage{xcolor}

\usepackage{graphicx}
\usepackage{soul}
\usepackage{wrapfig}

\usepackage{hyperref}
\hypersetup{
  colorlinks   = true,    % Colours links instead of ugly boxes
  urlcolor     = blue,    % Colour for external hyperlinks
  linkcolor    = blue,    % Colour of internal links
  citecolor    = red      % Colour of citations
}
\AtBeginDocument{%
  }

\definecolor{beautigreen}{RGB}{34,139,34}

\acrodef{RS}{recommender system}
\acrodef{MIND}{MIcrosoft News Dataset}

\newcommand{\header}[1]{\vspace*{1mm}\noindent\textbf{#1.}}

\AtBeginDocument{%
  }

\makeatletter
\newcommand{\ccell}[3][]{%
  \kern-\fboxsep
  \if\relax\detokenize{#1}\relax
    \expandafter\@firstoftwo
  \else
    \expandafter\@secondoftwo
  \fi
  {\colorbox{#2}}%
  {\colorbox[#1]{#2}}%
  {#3}\kern-\fboxsep
}
\makeatother
\definecolor{cellgray}{gray}{0.9}
\AtBeginDocument{\setlength{\cmidrulekern}{0.3em}}

\author{David Vos}
\orcid{0009-0003-8925-1585}
\affiliation{%
  \institution{University of Amsterdam}
  \city{Amsterdam}
  \country{The Netherlands}}
\email{d.j.a.vos@uva.nl}

\author{Andrew Yates}
\orcid{0000-0002-5970-880X}
\affiliation{%
  \institution{Johns Hopkins University}
  \city{Baltimore}
  \country{USA}
}
\email{andrew.yates@jhu.edu}

\author{Maarten de Rijke}
\orcid{0000-0002-1086-0202}
\affiliation{%
 \institution{University of Amsterdam}
 \city{Amsterdam}
 \country{The Netherlands}
}
\email{m.derijke@uva.nl}

\renewcommand{\shortauthors}{Vos et al.}

\copyrightyear{2026}
\acmYear{2026}
\setcopyright{cc}
\setcctype{by}
\acmConference[SIGIR-AP 2026]{Proceedings of the 2026 Annual International ACM SIGIR Conference on Research and Development in Information Retrieval in the Asia Pacific Region}{December 13--15, 2026}{Singapore, Singapore}
\acmBooktitle{Proceedings of the 2026 Annual International ACM SIGIR Conference on Research and Development in Information Retrieval in the Asia Pacific Region (SIGIR-AP 2026), December 13--15, 2026, Singapore, Singapore}
\acmDOI{10.1145/3848000.3850180}
\acmISBN{979-8-4007-3033-7/2026/12}


\begin{document}

\title[LSNR: Learned Sparse News Recommendations with Attentive User Modeling]{LSNR: Learned Sparse News Recommendations \\ with Attentive User Modeling}

%!TEX root = ../main.tex

\begin{abstract}
Current learned sparse approaches for recommender systems, such as STAR, represent a user by concatenating their interaction history into a single query string. For news, this creates two bottlenecks: concatenation collapses a user's distinct reading interests into one flat query, and, because news articles are long-form text, concatenating several news articles quickly overruns the context window of the language model used as backbone and forces quality-degrading truncation.

To address both bottlenecks, we introduce \emph{learned sparse news recommendations} (LSNR). LSNR encodes each article independently and constructs a sparse user vector by combining learned, attention-based weighting of the per-article sparse representations with a dense-projected expansion at the user level. Experiments on the MIND and EBNeRD datasets show that LSNR significantly outperforms STAR, with these gains holding even in short-history settings where truncation is not a factor. Compared to state-of-the-art dense recommenders, LSNR is competitive in standard settings and it outperforms them when the user history is short. An ablation of the user encoder confirms that learned attention-based aggregation outperforms simpler pooling alternatives. Beyond effectiveness, an additional advantage of LSNR is that it serves from an index an order of magnitude smaller than an uncompressed dense baseline. We release our code to facilitate reproducibility.\footnote{\url{https://github.com/davidvos/learned-sparse-news-recs}}
\end{abstract}

\maketitle

\section{Introduction}
\label{sec:intro}

News consumption is increasingly shifting toward digital platforms~\cite{vermeer-2020-online}. News recommender systems~\cite{karimi-2018-news, Raza2022-kr} help users navigate this landscape by providing personalized article suggestions. Modern approaches are dominated by deep learning methods that learn dense representations for users and news, achieving state-of-the-art performance on large-scale benchmarks~\cite{zhao-2025-revisiting, wu2021empowering}.

Learned sparse retrieval (LSR) has emerged as an alternative to dense retrieval. By using pre-trained language models to predict term importance and expand documents with latent terms, LSR produces sparse, lexical representations. Much like classical lexical models such as BM25, these representations can be indexed and retrieved with inverted-index machinery~\cite{lassance2024splade, tonellotto2018efficient, bruch2024efficient}, with every dimension mapping directly to a vocabulary term. In ad hoc retrieval, LSR methods have demonstrated strong in-domain and out-of-domain effectiveness~\cite{lassance2024splade,nguyen2026milco}.

Despite this success, the application of LSR to the news domain remains unexplored. The primary existing sparse recommender, STAR~\cite{tigunova2024star}, represents a user by concatenating their interaction history into a single query string. For news recommendation, this formulation is insufficient due to two critical bottlenecks. First, concatenation collapses a user's distinct reading interests into one undifferentiated query, leaving the model unable to weigh some articles as more relevant than others. Second, news articles are long-form text, so concatenating several of them quickly exceeds the context window of the underlying language model, forcing truncation that severely degrades recommendation quality. Both bottlenecks stem from the same root cause: STAR has no dedicated mechanism for representing a user from a collection of articles, and instead flattens the history into one document.

\header{Our approach}
We propose \emph{learned sparse news recommendations} (LSNR), the first learned sparse recommender designed for the news domain. Our central contribution is a user encoder that operates over independently encoded articles. Rather than flattening a user's history into a single string, LSNR encodes each article on its own and aggregates the resulting per-article sparse vectors with a learned, attention-based user encoder, fused with a dense-projected expansion at the user level. This design directly resolves both bottlenecks. It represents the user at the granularity of individual articles, learning which interactions matter rather than blending them together, and by encoding articles independently, it eliminates the context-window bottleneck. Because articles are encoded independently of the user, their sparse representations can be precomputed offline, leaving only user-side aggregation at inference time.

\header{Our findings}
On the MIND and EBNeRD benchmarks, LSNR significantly outperforms the leading sparse baseline STAR, improving NDCG@10 by 11\% on MIND. These gains hold even in short-history settings where truncation is not a factor. This demonstrates the importance of per-article modeling. Against state-of-the-art dense recommenders, the picture depends on the history length. In the standard full-history setting, LSNR performs very close to the strongest dense recommenders on MIND, within roughly half a percent on every metric, and trails them by 2 to 3\% on EBNeRD, while surpassing every dense baseline when the user history is short. 

This short-history advantage is especially consequential for news, where a large fraction of served users are new, infrequent, or sparsely tracked, so the regime in which LSNR overtakes dense models is one that news platforms encounter routinely.

Beyond effectiveness, the sparse formulation gives LSNR a compact footprint: it serves from an index that is an order of magnitude smaller than an uncompressed dense baseline. Because every dimension of the user vector maps to a vocabulary term, the vector can also be inspected directly.

\header{Our contributions}
\begin{enumerate*}[label=(\roman*)]
\item We introduce LSNR, the first learned sparse news recommender. We diagnose why STAR, the existing learned sparse recommender, fails to transfer to the news domain, and resolve it by replacing history concatenation with independent article encoding and a user encoder that aggregates the per-article representations with learned attention.
\item We show that LSNR significantly outperforms the STAR baseline on MIND and EBNeRD, maintaining this advantage even in short-history settings where truncation is absent.
\item We show that LSNR is competitive with dense models in the full-history setting, on par with the strongest dense recommenders on MIND and trailing them modestly on EBNeRD, while surpassing every dense baseline in short-history scenarios, the regime most characteristic of news.
\item An ablation of the user encoder shows that learned attention-based aggregation outperforms simpler pooling strategies, and that both pathways contribute to performance.
\item We release our code to facilitate reproducibility.\footnote{\label{ft:codelink}\url{https://github.com/davidvos/learned-sparse-news-recs}}
\end{enumerate*}
\section{Background}
\label{sec:background}

Two areas are particularly relevant to the work in this paper: news recommendation and learned sparse representations.

\subsection{News recommendation}
News recommendation poses unique challenges. Article content is short-lived, rapidly updated, and highly time-sensitive. This high turnover makes standard ID-based collaborative filtering approaches infeasible~\cite{karimi-2018-news}. Sequential recommenders such as SASRec~\cite{kang2018self} and its successors~\cite{Sun2019, Ding2022} are effective in domains with stable catalogs, but are generally considered suboptimal for news due to the rapid obsolescence of articles~\cite{wu2022news}. Even hybrid methods that combine sequential and content-based features are rarely applied to the news domain~\cite{li2023text}. Consequently, content-based neural news recommendation prevails. The standard architecture is a dual encoder that uses dense representations to model users and news~\cite{Raza2022-kr, Wu2023-ad, zhao-2025-revisiting}. Early approaches relied on sparse lexical features~\cite{Capelle2012-lw}, while pretrained embeddings shifted the focus to dense text encoders~\cite{Wang2018-uo, Wu2019-by, zhao-2025-revisiting, wu2021empowering}, sometimes enhanced with auxiliary signals~\cite{abdullah-2021-eliciting, Hu2020-lr, Wu2019-dz}. Large language models have also been explored~\cite{Gao2024-px, Liu2024-ds, li2023preliminary}, but their inference latency currently makes them infeasible for the retrieval stage focused on in this work~\cite{zhao-2025-revisiting}.

On the user side, dense dual encoders typically aggregate the history with attention, weighting interactions by their estimated importance. NRMS~\cite{Wu2019-by}, for instance, builds the user representation with multi-head self-attention over clicked articles. LSNR adopts a similar attention-based aggregation, but instead, the user representation is constructed in a sparse lexical space rather than a dense one. This yields a compact user representation whose dimensions correspond to vocabulary terms, whose properties we discuss next.

\subsection{Learned sparse representations}
Learned sparse representations (LSR) are commonly used in ad hoc retrieval to bridge the gap between exact lexical matching and semantic dense retrieval~\cite{dai2020contextaware, macavaney2020expansion, mallia2021learning, zhuang2021tilde, zhao2021sparta, lin2021briefnotes, lin2023dense, doshi2024mistral, zhuang2021fast}. State-of-the-art LSR methods, e.g., SPLADE~\cite{formal-2021-splade, lassance2024splade}, use masked language modeling heads to predict term importance and to expand documents with relevant latent terms~\cite{nguyen2023unified}. These methods allow for flexible efficiency and effectiveness trade-offs via techniques such as block-max pruning and quantized inverted indices~\cite{mallia2024faster, mackenzie2022accelerating, bruch2024efficient, carlson2025superblock}. Because every dimension corresponds to a vocabulary term, these representations are often described as interpretable, though recent analyses show that SPLADE can route ranking signals through terms only loosely related to the input, or even through irrelevant tokens~\cite{mackenzie2023exploring}.

News recommendation differs fundamentally from ad hoc retrieval. The input is not a text string but a history of previously read articles, from which a user representation must be constructed. The pioneering learned sparse recommender, STAR~\cite{tigunova2024star}, adapts LSR by concatenating the interaction history into a single text string. As discussed in Section~\ref{sec:intro}, this is ill-suited to news: it collapses a user's distinct interests into one undifferentiated representation and quickly exceeds the language model's context window, forcing quality-degrading truncation. We instead encode each article independently and aggregate the resulting sparse representations with a learned user encoder, representing the user at the granularity of individual articles rather than as a single concatenated string.

\usetikzlibrary{positioning, arrows.meta, calc}

\begin{figure}[t!]
    \centering
    \definecolor{proBlue}{RGB}{26, 95, 122}
    \definecolor{proOrange}{RGB}{242, 100, 25}
    \definecolor{proGreen}{RGB}{79, 119, 45}
    \definecolor{proPurple}{RGB}{120, 60, 140}

    \resizebox{\columnwidth}{!}{%
    \begin{tikzpicture}[
        x=1mm, y=-1mm,
        font=\sffamily\footnotesize,
        >=Stealth,
        box/.style={rectangle, draw, thick, rounded corners=2pt, align=center,
                     text=black!90, inner sep=3pt, minimum height=9mm},
        hist/.style={box, draw=proBlue!80!black, fill=proBlue!12},
        dense/.style={box, draw=proBlue!70!cyan!80!black, fill=proBlue!70!cyan!12},
        proj/.style={box, draw=proPurple!80!black, fill=proPurple!10},
        cand/.style={box, draw=proOrange!90!black, fill=proOrange!12},
        scr/.style={box, draw=proGreen!80!black, fill=proGreen!12, rounded corners=1pt},
        arr/.style={->, thick, black!55, shorten >=1pt, shorten <=1pt},
        darr/.style={->, thick, black!45, dashed, shorten >=1pt, shorten <=1pt},
        shared_link/.style={<->, thick, densely dashed, black!50},
    ]

        % === Layer 1: Inputs ===
        \node[hist, minimum width=22mm] (in_h) at (16,0)
            {\textbf{User history}\\\scriptsize $\mathbf{A}_1 \dots \mathbf{A}_N$};
        \node[cand, minimum width=22mm] (in_c) at (52,0)
            {\textbf{Candidate}\\\scriptsize $\mathbf{A}_c$};

        % === Layer 2: Encoders ===
        \node[hist, minimum width=22mm] (enc_h) at (16,14)
            {\textbf{News encoder}\\\scriptsize SPLADE};
        \node[cand, minimum width=22mm] (enc_c) at (52,14)
            {\textbf{News encoder}\\\scriptsize SPLADE};

        % === Layer 3: Representations ===
        \node[hist, minimum width=15mm] (s_h) at (8,28)
            {\textbf{Sparse}\\\scriptsize $\mathbf{s}_i$};
        \node[dense, minimum width=15mm] (d_h) at (24,28)
            {\textbf{Dense}\\\scriptsize $\mathbf{d}_i$};
        \node[cand, minimum width=15mm] (s_c) at (52,28)
            {\textbf{Sparse}\\\scriptsize $\mathbf{s}_c$};

        % === Layer 4: Attention ===
        \node[dense, minimum width=15mm] (attn) at (24,42)
            {\textbf{Attention}\\\scriptsize $\alpha_i$};

        % === Layer 5: Pathways ===
        \node[hist, minimum width=16mm, minimum height=10mm] (u_h) at (8,56)
            {$\mathbf{u}_{\mathrm{lex}}$\\\scriptsize $\sum_i \alpha_i \mathbf{s}_i$};
        \node[dense, minimum width=20mm, minimum height=10mm] (u_p) at (28,56)
            {$\mathbf{u}_{\text{sem}}$\\\scriptsize SPLADE($\mathbf{W}\mathbf{u}_{\mathrm{d}}$)};

        % === Layer 6: User vector & Score ===
        \node[hist, minimum width=24mm, minimum height=10mm] (u) at (18,70)
            {$\mathbf{u} = \mathbf{u}_{\mathrm{lex}} + \mathbf{u}_{\mathrm{sem}}$};
        \node[scr, minimum width=20mm, minimum height=10mm] (sc) at (52,70)
            {\textbf{Score}\\\scriptsize $\mathbf{u}^\top\!\mathbf{s}_c$};

        % === Edges ===
        % Inputs -> Encoders
        \draw[arr] (in_h) -- (enc_h);
        \draw[arr] (in_c) -- (enc_c);

        % Shared weights
        \draw[shared_link] (enc_h.east) -- (enc_c.west)
            node[midway, above, font=\sffamily\scriptsize\bfseries,
                 text=black!65] {shared};

        % Encoder -> Representations
        \draw[arr] (enc_h.south) -- (s_h.north);
        \draw[arr] (enc_h.south) -- (d_h.north);
        \draw[arr] (enc_c.south) -- (s_c.north);

        % Dense -> Attention
        \draw[arr] (d_h) -- (attn);

                % Sparse history -> u_hist (straight down)
        \draw[arr] (s_h) -- (u_h);

        % Attention -> pathways (dashed, both from bottom-center)
        \draw[darr] (attn.south) -- (u_h.north);
        \draw[darr] (attn.south) -- (u_p.north);

        % Pathways -> User vector
        \draw[arr] (u_h) -- (u.north -| u_h);
        \draw[arr] (u_p) -- (u.north -| u_p);

        % Candidate sparse -> Score (straight vertical)
        \draw[arr] (s_c) -- (sc);

        % User vector -> Score
        \draw[arr] (u.east) -- (sc.west);

    \end{tikzpicture}
    }%

    \vspace{2mm}
\caption{LSNR architecture. A shared news encoder maps each article to a sparse representation $\mathbf{s}$ and, for history items, a dense representation $\mathbf{d}$. Additive attention over the dense history vectors produces weights $\alpha_i$. The sparse user vector $\mathbf{u}$ is the sum of two pathways: a lexical pathway $\mathbf{u}_{\mathrm{lex}}$ that aggregates per-article sparse vectors using attention weights, and a semantic pathway $\mathbf{u}_{\mathrm{sem}}$ that applies a SPLADE projection to the attention-weighted dense user representation. A candidate article is scored using the sparse inner product $\mathbf{u}^\top \mathbf{s}_c$.}
\label{fig:architecture}

\end{figure}
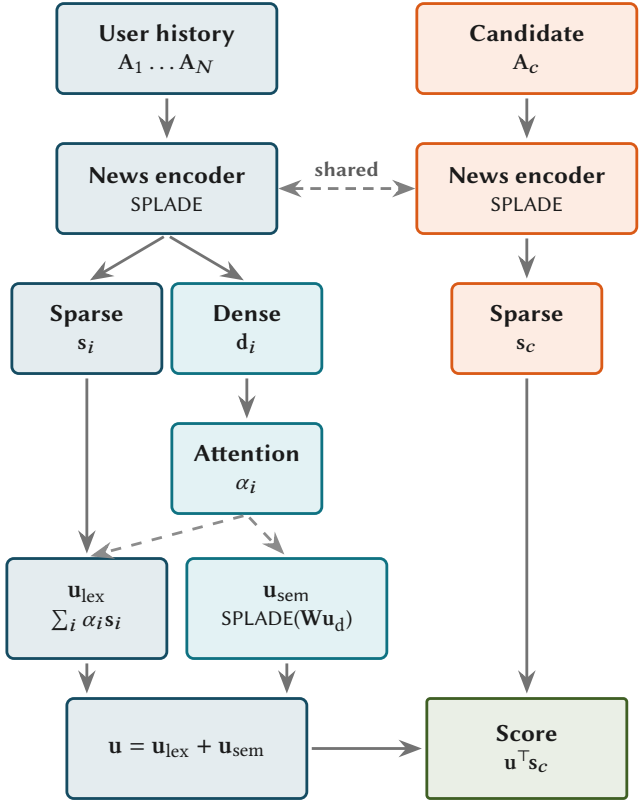

\section{Methodology}
\label{sec:method}

We propose learned sparse news recommendations (LSNR). See Figure~\ref{fig:architecture}. The news encoder independently encodes each news article as a sparse lexical representation, based on SPLADE~\cite{formal-2021-splade}, and produces an auxiliary dense representation that guides aggregation in the user encoder. The latter constructs a sparse user vector via an attention-weighted aggregation of the per-article sparse representations and a dense-projected expansion at the user level. The news encoder is applied to history and candidate articles, so candidate representations can be precomputed once and indexed offline.

\subsection{News encoder}
Given a news article consisting of tokens $\{w_1, \dots, w_L\}$, we use a pre-trained language model (e.g., BERT) to obtain contextualized hidden states $\mathbf{H} \in \mathbb{R}^{L \times d}$, where $\mathbf{h}_i$ denotes the hidden state of token $i$. Following common practice for adapting pre-trained backbones to news recommendation tasks \cite{wu2021empowering}, we pass these hidden states through an additional trainable transformer layer before projecting to the sparse vocabulary space. We project each token to the sparse vocabulary space $\mathcal{V}$ using the SPLADE log-saturation activation. The importance $w_{ij}$ of vocabulary term $j$ for token $i$ is computed as:
\begin{equation}
    w_{ij} = \log\bigl(1 + \operatorname{ReLU}(\mathbf{W}_{vocab}\, \mathbf{h}_i)\bigr)_j,
    \label{eq:splade}
\end{equation}
where $\mathbf{W}_{vocab}$ is initialized from, and tied to, the input word embeddings of the backbone, following standard SPLADE practice~\cite{formal-2021-splade}. The final sparse article representation $\mathbf{s} \in \mathbb{R}^{|\mathcal{V}|}$ is obtained via max-pooling over the sequence length: $\mathbf{s}_j = \max_{i} w_{ij}$. 
Simultaneously, we extract a dense representation $\mathbf{d} \in \mathbb{R}^d$ via mean-pooling over the transformer-layer outputs. The dense representation is used only for history articles, where it guides aggregation in the user encoder. Candidate articles are represented by their sparse vector $\mathbf{s}$ alone (Section~\ref{sec:scoring}). 
Table~\ref{tab:example} illustrates a learned representation, showing how LSNR captures both exact keywords and semantic expansions (e.g., \textit{shop} $\rightarrow$ \textit{buying}) while avoiding the semantic collapse that can be observed in sparse methods~\cite{nguyen2024multimodal}.
\looseness=-1

\begin{table}[t]
\centering
\setlength{\tabcolsep}{3pt}
\caption{Title representation and LSNR representation of MIND article N88753. Bold tokens indicate exact lexical matches with the title text, while others are semantic expansions.}
\begin{tabular}{@{}p{0.12\linewidth} p{0.86\linewidth}@{}}
\toprule
\textbf{Title} & The Brands Queen Elizabeth, Prince Charles, and Prince Philip Swear By. Shop the notebooks, jackets, and more that the royals can't live without. \\
\midrule
\textbf{Top \mbox{tokens}} &
\textbf{royal}, liberal, marine, conservative, imperial, victoria, fan, wedding, fans, diplomatic, buying, york, \textbf{brand}, \textbf{elizabeth}, setting \\
\bottomrule
\end{tabular}
\label{tab:example}
\end{table}

\subsection{User encoder}
\label{sec:user-encoder}
\looseness=1
The main contribution of LSNR is to produce a sparse user vector from a set of independently encoded articles. The encoder uses dense representations to determine which articles matter, and combines two  pathways for content: a direct aggregation of the per-article sparse representations and a SPLADE projection from the aggregated dense vector.

Given a user history of $N$ articles with dense representations $\mathbf{D} \in \mathbb{R}^{N \times d}$ and sparse representations $\mathbf{S} \in \mathbb{R}^{N \times |\mathcal{V}|}$, we first contextualize the dense vectors using a single transformer layer~\cite{vaswani2017attention}. This allows each article to be represented in the context of the others, yielding contextualized dense vectors $\{\tilde{\mathbf{d}}_i\}_{i=1}^N$. We omit positional encodings, as we find empirically that they do not improve performance (Section~\ref{sec:ablation}). This is consistent with prior findings that strict temporal order is a weak signal for news recommendation~\cite{wu2022news}.

To determine the importance of each interaction, we apply an additive attention mechanism over the contextualized dense vectors. For each history item, we compute an attention score and normalize it across the history:
\begin{equation}
    e_i = \mathbf{q}^\top \tanh(\mathbf{W}_a \tilde{\mathbf{d}}_i + \mathbf{b}_a),
    \qquad
    \alpha_i = \frac{\exp(e_i)}{\sum_{j=1}^N \exp(e_j)},
    \label{eq:attn}
\end{equation}
where $\mathbf{W}_a$, $\mathbf{b}_a$, and the attention-query vector $\mathbf{q}$ are learned parameters.

\paragraph{Two-pathway aggregation.}
We construct the sparse user vector $\mathbf{u} \in \mathbb{R}^{|\mathcal{V}|}$ from two complementary pathways: a lexical pathway that aggregates the per-article sparse representations directly, and a semantic pathway that projects the aggregated dense representation into sparse space. The lexical pathway combines the per-article sparse vectors using the attention weights
\begin{equation}
    \mathbf{u}_{\text{lex}} = \sum_{i=1}^N \alpha_i \, \mathbf{s}_i,
    \label{eq:u-lex}
\end{equation}
producing a vector whose non-zero entries are inherited from the SPLADE expansions of individual history articles. The semantic pathway applies the SPLADE activation to a projection from the attention-weighted dense user representation
\begin{equation}
    \mathbf{u}_{\text{sem}} = \log\Bigl(1 + \operatorname{ReLU}\bigl(\mathbf{W}_{vocab} \textstyle\sum_{i=1}^N \alpha_i \, \tilde{\mathbf{d}}_i\bigr)\Bigr),
    \label{eq:u-sem}
\end{equation}
where $\mathbf{W}_{vocab}$ is the same vocabulary projection used in the news encoder (Eq.~\ref{eq:splade}). The final user vector is the sum of the two pathways:
\begin{equation}
    \mathbf{u} = \mathbf{u}_{\text{lex}} + \mathbf{u}_{\text{sem}}.
    \label{eq:u-final}
\end{equation}

\begin{table*}[t!]
\centering
%\small
\caption{Sparse user vectors produced by LSNR's user encoder on three MIND validation users. Each vector shows the top-20 active tokens with their activation weights. \textbf{Bold} tokens are exact lexical matches with the user's clicked article texts. The remaining tokens are semantic expansions (e.g., \textit{talent, artists, music} for an entertainment reader and \textit{kitchen, coffee, pizza} for a food reader), illustrating how the learned representation captures interests beyond the literal click history. These tokens reflect the combined contributions of both pathways in the hybrid user encoder. These users were selected from validation for having clear single-category dominance.}
\label{tab:user_profiles_mind}
\setlength{\tabcolsep}{4pt}
\renewcommand{\arraystretch}{1.15}
\begin{tabular}{@{}>{\bfseries}l p{0.92\linewidth}@{}}
\toprule
\multicolumn{2}{@{}l}{\textbf{User A} \hfill idx 737 \quad $\bullet$ \quad 34 articles \quad $\bullet$ \quad 50\% Finance} \\
\midrule
Clicks &
\textit{America's largest auto retailer sells customers defective used cars, report says}.
\textit{My husband and I saved \$36{,}000 in 2 years to quit our jobs and travel the world}.
\textit{`SOS' picked up on CCTV saves Aussie woman lost in bush}.
\textit{Clear your browser before booking a flight to get a better deal}. \\
\addlinespace[2pt]
Profile &
bank\,(0.16), \textbf{rich}\,(0.16), buying\,(0.15), architect\,(0.14), \textbf{china}\,(0.12), \textbf{my}\,(0.11), banking\,(0.10), jet\,(0.09), \textbf{animal}\,(0.07), me\,(0.07), \textbf{forbes}\,(0.07), demolished\,(0.07), picture\,(0.06), stock\,(0.06), cat\,(0.06), american\,(0.06), \textbf{tour}\,(0.06), \textbf{cars}\,(0.05), \textbf{on}\,(0.05), poker\,(0.05) \\
\midrule
\multicolumn{2}{@{}l}{\textbf{User B} \hfill idx 1252 \quad $\bullet$ \quad 37 articles \quad $\bullet$ \quad 57\% Entertainment} \\
\midrule
Clicks &
\textit{Chrissy Teigen Predicts The Office's Jim and Pam Are Divorced Now}.
\textit{Millie Bobby Brown Reveals Her Style Inspo, Celebrity Crush \& More}.
\textit{`Wheel of Fortune' Guest Delivers Hilarious, Off the Rails Introduction}.
\textit{Daughter Arrested After Calling 911 When Mom Collapses at Concert}. \\
\addlinespace[2pt]
Profile &
\textbf{stars}\,(0.31), talent\,(0.23), artists\,(0.22), picture\,(0.14), mall\,(0.12), \textbf{beauty}\,(0.11), rich\,(0.11), city\,(0.11), web\,(0.11), styles\,(0.10), ball\,(0.08), studios\,(0.08), revolution\,(0.07), music\,(0.07), urban\,(0.06), designed\,(0.06), children\,(0.05), design\,(0.05), photographer\,(0.05), cool\,(0.05) \\
\midrule
\multicolumn{2}{@{}l}{\textbf{User C} \hfill idx 94 \quad $\bullet$ \quad 24 articles \quad $\bullet$ \quad 54\% Food \& Drink} \\
\midrule
Clicks &
\textit{Lay's newest flavor is grilled cheese and tomato soup}.
\textit{Burger King's new `Ghost Whopper' is here to haunt your Halloween order}.
\textit{Kelly Ripa jokes son Michael is experiencing `extreme poverty' in Brooklyn}.
\textit{Popeyes quietly launched a new pumpkin pie}. \\
\addlinespace[2pt]
Profile &
librarian\,(0.24), kitchen\,(0.22), demolished\,(0.16), pupils\,(0.15), coffee\,(0.13), tour\,(0.11), \textbf{chicken}\,(0.11), removed\,(0.11), beach\,(0.10), sulfur\,(0.09), pizza\,(0.08), studios\,(0.08), \textbf{better}\,(0.08), long\,(0.07), \textbf{pie}\,(0.07), intervention\,(0.07), short\,(0.07), mall\,(0.06), \textbf{me}\,(0.06), rolled\,(0.06) \\
\bottomrule
\end{tabular}
\end{table*}

\noindent%
The two pathways play complementary roles. $\mathbf{u}_{\text{lex}}$ carries a direct lexical signal: each non-zero entry can be traced to a weighted combination of the SPLADE expansions of specific history articles. $\mathbf{u}_{\text{sem}}$ allows the model to activate terms that are not present in any single article's expansion but that the aggregate dense interest profile points toward, providing user-level semantic expansion beyond what per-article expansion can capture. Both pathways depend on the same attention weights $\alpha_i$ and the same vocabulary projection $\mathbf{W}_{vocab}$, so the hybrid user encoder introduces no additional learnable parameters beyond what a single-pathway variant would require. We ablate the contribution of each pathway, and of the aggregation function within $\mathbf{u}_{\text{lex}}$, in Section~\ref{sec:ablation}.  Because every dimension of $\mathbf{u}$ corresponds to a vocabulary term, the user vector can be read at the term level, as illustrated in \autoref{tab:user_profiles_mind} (though SPLADE weights need not be strictly semantic~\cite{mackenzie2023exploring}).

Overall, the user encoder defines an aggregation that decomposes into three components: 
\begin{enumerate*}[label=(\roman*)]
\item cross-article contextualization of the dense vectors, 
\item learned importance weighting via additive attention, and 
\item a two-pathway sparse construction that combines direct sparse aggregation with dense-projected expansion. 
\end{enumerate*}
As parameter-free alternatives, we also experiment with temporal-decay weighted, mean, and max pooling over $\{\mathbf{s}_i\}$ (replacing the $\mathbf{u}_{\text{lex}}$ pathway while keeping $\mathbf{u}_{\text{sem}}$ unchanged) in Section~\ref{sec:ablation}.

\vspace{0mm}
\subsection{Scoring}
\label{sec:scoring}
At retrieval time, a candidate article is scored against the user vector using the inner product of their sparse representations, $\operatorname{score}(\mathbf{u}, c) = \mathbf{u}^\top \mathbf{s}_c$. Because both vectors are sparse, this is computed efficiently over an inverted index. Furthermore, because candidate representations $\mathbf{s}_c$ are user-independent, they are precomputed at indexing time (Section~\ref{sec:efficiency}).

\subsection{Regularization}
To regularize the sparsity of LSNR representations, we apply an $\mathcal{L}_{\text{FLOPS}}$ regularizer during training~\cite{lassance2024splade}, with distinct coefficients $\lambda_n$ for candidate news articles and $\lambda_u$ for the final aggregated user vector $\mathbf{u}$, which combines both pathways via Eq.~\ref{eq:u-final}. We fix $\lambda_n = 0.005$ and $\lambda_u = 0.0005$, tuned on the validation set. LSNR news representations have an average of 22.8 (MIND) and 31.6 (EBNeRD) non-zero dimensions, and user vectors average 722 (MIND) and 563 (EBNeRD) non-zero dimensions before retrieval-time pruning. User vectors are naturally denser than news vectors, but still use less than 3\% of the vocabulary space of the underlying BERT-based language models. We return to the implications of this sparsity for retrieval efficiency in Section~\ref{sec:efficiency}.

\section{Experimental setup}
\label{sec:setup}

We describe the datasets that we use for evaluation, the baselines that we compare against, optimization details, and our evaluation protocols.

\subsection{Datasets}
We evaluate LSNR on two public news recommendation benchmarks: the Microsoft News Dataset (MIND)~\cite{Wu2020-rg} in English, and the Ekstra Bladet News Recommendation Dataset (EBNeRD)~\cite{Kruse2024-st} in Danish. Following standard practice~\cite{kannen-etal-2024-efficient, zhao-2025-revisiting}, we use the ``small'' versions of both datasets. Since the official test sets lack public labels, we adopt the convention of using the official validation split for testing and sample the last 10\% of the training split for hyperparameter tuning~\cite{zhao-2025-revisiting}. \autoref{tab:dataset-stats} summarizes the statistics. The two datasets are complementary: MIND consists of short articles (title and abstract) with moderate user histories, whereas EBNeRD features substantially longer articles (including full body text) and longer histories.

\begin{table}[t]
\centering
\caption{MIND and EBNeRD dataset statistics. \#Users, \#Art., and \#Int. give the number of users, articles, and interactions respectively. $\overline{H}$ is average user history length and $\overline{W}$ is average words per article.}
\label{tab:dataset-stats}
\setlength{\tabcolsep}{5pt}
\begin{tabular}{ll rrrr c}
\toprule
\textbf{Dataset} & \textbf{Split} & \textbf{\#Users} & \textbf{\#Art.} & \textbf{\#Int.} & $\overline{\textbf{H}}$ & $\overline{\textbf{W}}$ \\
\midrule
\multirow{3}{*}{MIND}
  & Train & 48{,}052 & 49{,}476 & 212{,}481 & 39.2 & \multirow{3}{*}{45.0\textsuperscript{a}} \\
  & Val   & 11{,}995 & 51{,}282 &  15{,}697 & 33.0 & \\
  & Test  & 50{,}000 & 65{,}238 &  73{,}152 & 32.3 & \\
\midrule
\multirow{3}{*}{EBNeRD}
  & Train & 14{,}834 &  7{,}317 & 210{,}402 & 47.2 & \multirow{3}{*}{465.9\textsuperscript{b}} \\
  & Val   &  6{,}637 &  7{,}527 &  23{,}289 & 47.4 & \\
  & Test  & 15{,}342 &  9{,}985 & 244{,}647 & 46.5 & \\
\bottomrule
\end{tabular}
\par\smallskip
\begin{minipage}{\columnwidth}
\scriptsize
\textsuperscript{a} Title + Abstract (Title: 10.8, Abstract: 34.2 words).\\
\textsuperscript{b} Title + Abstract + Body (Title: 6.6, Abstract: 19.5, Body: 439.8 words).
\end{minipage}
\end{table}

\subsection{Baseline methods}
To validate the effectiveness of LSNR, we compare it against both sparse and dense content-based recommender systems. We first test our hypothesis that learned, attention-based aggregation is superior to concatenation by comparing against \textbf{BM25} and \textbf{STAR}~\cite{tigunova2024star}, the state-of-the-art learned sparse recommender. We adapt STAR for the news domain by concatenating the most recent interactions, as we found recency-based selection more effective than the original IDF-based selection. We concatenate historical interactions into STAR's input until the 512-token context window is filled. For the user-encoder ablation, we additionally implement several aggregation variants and pathway configurations in place of our default hybrid user encoder, described in Section~\ref{sec:ablation}.

We benchmark LSNR against three leading dense architectures, \textbf{NRMS}~\cite{Wu2019-by}, \textbf{NAML}~\cite{Wu2019-to}, and \textbf{LSTUR}~\cite{An2019-cf}, each implemented with a pre-trained language model backbone to align with modern standards~\cite{wu2021empowering, zhao-2025-revisiting}. Recent work has shown that BERT-based variants of these models perform on par with much larger LLM-based encoders~\cite{zhao-2025-revisiting}, justifying our use of BERT-scale backbones. LSNR, STAR, and NRMS use only article texts. NAML and LSTUR additionally include article categories and user IDs. We exclude generative LLMs due to their prohibitive inference costs at the retrieval stage~\cite{zhao-2025-revisiting}.

\vspace{0mm}
\subsection{Optimization details}
We initialize English models with the \textit{bert-base-uncased} checkpoint and Danish models with the \textit{Maltehb/danish-bert-botxo} checkpoint. For STAR on MIND, we use the \textit{naver/splade-cocondenser-ensembledistil} checkpoint to match the original implementation. As no Danish SPLADE checkpoint is available, we train STAR on EBNeRD from the same \textit{danish-bert-botxo} backbone as LSNR, so both sparse models start from identical initialization. On MIND, this initialization asymmetry favors STAR, which begins from a retrieval-pretrained checkpoint while LSNR begins from vanilla BERT. The additional transformer layer in the news encoder and the contextualization layer in the user encoder each use a single layer with 16 attention heads. Following~\cite{zhao-2025-revisiting} in freezing the lower backbone layers to balance performance and efficiency, we freeze the first 9 of the 12 backbone layers across all methods. We optimize using AdamW on a single NVIDIA H100 GPU, grid-searching the learning rate over $\{10^{-6}, 10^{-5}, 10^{-4}\}$ with early stopping on validation NDCG@10.

We tune the negative sampling strategy per model to give each its strongest configuration. For dense models, we sample 4 negatives from the candidate pool. Increasing this yielded no improvement, consistent with prior work~\cite{wu2021empowering}. For sparse models (LSNR and STAR), we sample 16 negatives, as we found learned sparse models benefit from a larger negative pool.

\subsection{Evaluation protocols}
We structure our evaluation in five stages, each targeting a distinct question. We report NDCG@10, NDCG@5, and MRR, and assess statistical significance using a two-tailed paired t-test over the per-user metric distributions, where $^\ast$ denotes a significant improvement over the second-best method ($p < 0.01$). Following the standard MIND and EBNeRD protocol, all NDCG and MRR metrics throughout the paper are computed by ranking the candidate articles recorded in each impression log, treating clicked articles as positives and co-displayed non-clicked articles as negatives, and averaging over impressions. The full-catalog retrieval used for the efficiency measurements in Section~\ref{sec:efficiency} applies only to index footprint and latency, not to accuracy.

\header{Sparse baselines}
We first assess recommendation performance on the full test set against the sparse baselines BM25 and STAR. This stage tests our core hypothesis that learned aggregation over independently encoded articles outperforms STAR's concatenation.

\header{Short-history robustness}
A natural concern is that LSNR may outperform STAR only because STAR is forced to truncate long histories. To isolate the effect of our aggregation mechanism from the avoidance of truncation, we simulate short-history users by training on the complete training set but truncating test-set histories to the $N$ most recent interactions, $N \in \{1, \dots, 5\}$. At small $N$, and especially at $N=1$, STAR processes the exact same input as LSNR with no truncation, so any remaining gap is attributable to how each model represents the user rather than to lost input.

\header{Dense baselines}
We benchmark LSNR against the state-of-the-art dense recommenders NRMS, NAML, and LSTUR, on both the full dataset and the short-history setting, to assess whether learned sparse representations can match dense embeddings.

\header{Efficiency}
We report the index footprint of LSNR against the dense NRMS baseline. Because LSNR user vectors are sparse, we also examine the effect of retrieval-time pruning: we sweep a threshold $K \in \{5, \dots, 100\}$ that retains only the top-$K$ highest-scoring terms of each user vector, and trace recommendation accuracy against user-vector sparsity (non-zero entries, NNZ). We report user-vector and document sparsity and on-disk index size, comparing LSNR against the uncompressed dense NRMS baseline on identical hardware, and report end-to-end retrieval latency.

\header{User aggregation}
Finally, to justify the design choices in the user encoder, we run two complementary ablations. The first isolates the contribution of each pathway in the hybrid construction (Eq.~\ref{eq:u-final}), training variants that disable $\mathbf{u}_{\text{lex}}$, disable $\mathbf{u}_{\text{sem}}$, or include both. The second varies the aggregation function within $\mathbf{u}_{\text{lex}}$, comparing our additive attention against four alternatives: attention with positional encodings, mean pooling, max pooling, and a recency-weighted temporal decay. A comparison against the parameter-free pooling variants isolates the contribution of learned weighting, while the positional-encoding and temporal-decay variants test whether interaction order carries signal beyond article content. In all variants except the pathway ablation, the $\mathbf{u}_{\text{sem}}$ pathway is held constant. We report NDCG@10 in both the full-history and short-history ($N=1$) settings.

\section{Results}

We evaluate the performance of LSNR across several key dimensions: effectiveness against sparse and dense baselines, retrieval efficiency, and the contribution of its individual components. We show that LSNR consistently outperforms sparse models and remains highly competitive with state-of-the-art dense recommenders while offering a more compact index.

\subsection{Benchmarking LSNR against sparse baselines}
\label{sec:results-sparse}
We first evaluate LSNR against sparse baselines to test our core hypothesis that learned aggregation over independently encoded articles is superior to naive history concatenation. \autoref{tab:sparse_results_full} reports the performance on the full MIND and EBNeRD test sets.

\begin{table}[h]
\caption{Performance of LSNR compared to BM25 and learned sparse baseline STAR across the full MIND and EBNeRD benchmarks. \textbf{Bold} denotes the highest score, $^\ast$ denotes a significant improvement over the second-best method ($p < 0.01$).}
\label{tab:sparse_results_full}
\setlength{\tabcolsep}{2.7pt}
\centering
\begin{tabular}{l ccc ccc}
\toprule
 & \multicolumn{3}{c}{\bf MIND} & \multicolumn{3}{c}{\bf EBNeRD} \\
\cmidrule(lr){2-4} \cmidrule(lr){5-7}
\textbf{Method} & N@10 & N@5 & MRR & N@10 & N@5 & MRR \\
\midrule
BM25 & 0.3142 & 0.2541 & 0.2480 & 0.4639 & 0.3868 & 0.3503 \\
STAR & 0.3815 & 0.3191 & 0.2973 & 0.4704 & 0.3925 & 0.3459 \\
\midrule
\textbf{LSNR (Ours)} & \textbf{0.4234}\rlap{$^\ast$} & \textbf{0.3620}\rlap{$^\ast$} & \textbf{0.3310}\rlap{$^\ast$} & \textbf{0.5200}\rlap{$^\ast$} & \textbf{0.4547}\rlap{$^\ast$} & \textbf{0.4027}\rlap{$^\ast$} \\
\bottomrule
\end{tabular}
\end{table}

\noindent%
Across both datasets, LSNR significantly outperforms both BM25 and the state-of-the-art sparse baseline STAR. On MIND, LSNR achieves a gain of 11\% over STAR in NDCG@10, and over 10\% on EBNeRD. The contrast is sharpest on EBNeRD, where STAR's advantage over the unsupervised BM25 baseline nearly vanishes: STAR edges BM25 by less than a point on NDCG@10 (0.4704 vs.\ 0.4639) and trails it on MRR (0.3459 vs.\ 0.3503). On these long-article histories (\autoref{tab:dataset-stats}), concatenation improves STAR almost nothing over a purely lexical baseline, consistent with the context-window bottlenecks it suffers. These results empirically demonstrate that replacing concatenation with independent article encoding and learned, two-pathway aggregation allows LSNR to better capture distinct user interests at the per-article granularity.

\subsection{LSNR in short-history settings}
A potential counter-argument against the main finding from \autoref{sec:results-sparse} is that LSNR outperforms STAR only because STAR is forced to truncate long histories. To control for this, we evaluate both models in a simulated short-history setting ($N=1$), where STAR suffers no truncation and processes the exact same input as LSNR.

As reported in \autoref{tab:short_history_sparse}, LSNR consistently outperforms STAR even when history is limited to a single interaction, by 3.6\% in NDCG@10 on MIND and by over 7\% on EBNeRD, with both gains statistically significant. We attribute this to a difference in training signal. STAR optimizes the representation of a single concatenated string per user, whereas LSNR optimizes each article's representation against per-user-vector gradients, producing per-article representations that generalize better when interaction data is scarce.

\begin{table}[t]
\caption{Short-history ($N=1$) performance of LSNR compared to BM25 and learned sparse baseline STAR. \textbf{Bold} denotes the highest score, $^\ast$ denotes a significant improvement over the second-best method ($p < 0.01$).}
\label{tab:short_history_sparse}
%\small
\setlength{\tabcolsep}{2.7pt}
\centering
\begin{tabular}{l ccc ccc}
\toprule
 & \multicolumn{3}{c}{\bf MIND ($N=1$)} & \multicolumn{3}{c}{\bf EBNeRD ($N=1$)} \\
\cmidrule(lr){2-4} \cmidrule(lr){5-7}
\textbf{Method} & N@10 & N@5 & MRR & N@10 & N@5 & MRR \\
\midrule
BM25 & 0.3109 & 0.2505 & 0.2445 & 0.4359 & 0.3519 & 0.3203 \\
STAR & 0.3500 & 0.2880 & 0.2742 & 0.4656 & 0.3866 & 0.3408 \\
\midrule
\textbf{LSNR (Ours)} & \textbf{0.3626}\rlap{$^\ast$} & \textbf{0.3015}\rlap{$^\ast$} & \textbf{0.2819}\rlap{$^\ast$} & \textbf{0.5022}\rlap{$^\ast$} & \textbf{0.4317}\rlap{$^\ast$} & \textbf{0.3824}\rlap{$^\ast$} \\
\bottomrule
\end{tabular}
\end{table}

\begin{figure}[h]
    \centering
    \includegraphics[width=0.49\linewidth]{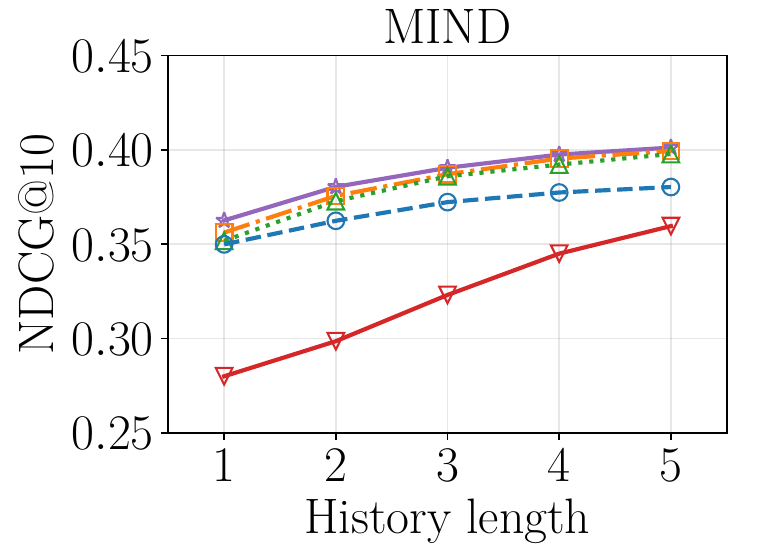}
    \includegraphics[width=0.49\linewidth]{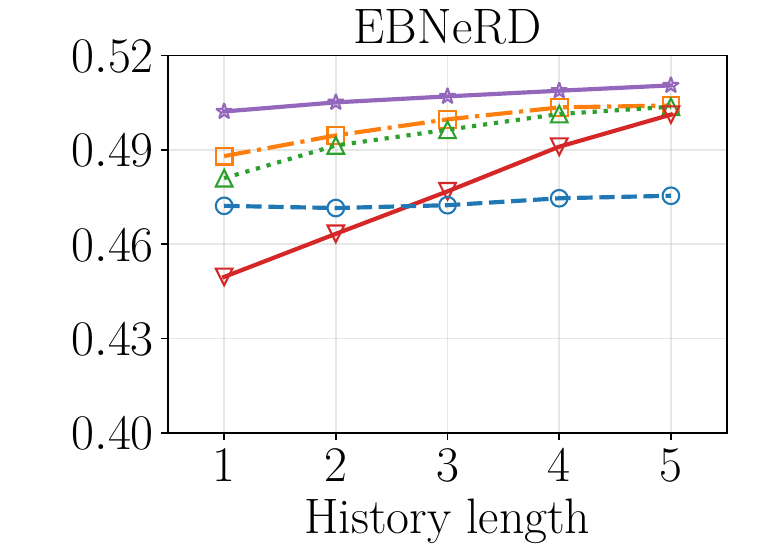}
    \includegraphics[width=\linewidth]{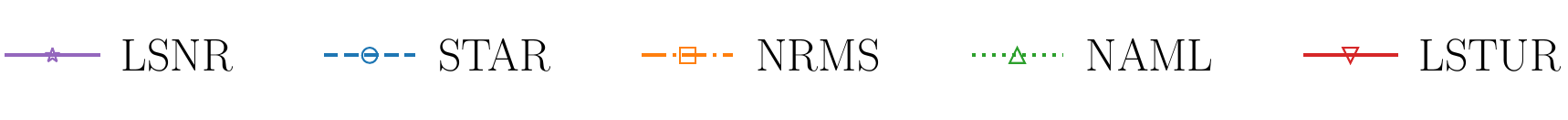}
    \caption{NDCG@10 of LSNR, STAR, and the dense baselines as user history length increases from 1 to 5 on MIND (left) and EBNeRD (right).}
    \label{fig:history_length_performance}
\end{figure}

\autoref{fig:history_length_performance} visualizes performance as history length grows from 1 to 5. LSNR improves rapidly, whereas STAR yields diminishing returns. This divergence confirms that concatenating text is an ineffective mechanism for aggregating user interactions in news recommendation, compared to the learned attention-based weighting employed by LSNR.

\subsection{Comparing LSNR with dense baselines}

Next, we benchmark LSNR against state-of-the-art dense recommenders (pre-trained language model variants of NRMS, NAML and LSTUR) to assess whether learned sparse representations can match dense embeddings. \autoref{tab:dense_comparison} presents the performance on both full and short-history settings.

\begin{table*}[t!]
\caption{Comparison of LSNR against state-of-the-art dense news recommenders. \textbf{Bold} denotes the highest score, $^\ast$ denotes a significant improvement over the second-best method ($p < 0.01$).}
\label{tab:dense_comparison}
\centering
\begin{tabular}{l ccc ccc ccc ccc}
\toprule
 & \multicolumn{6}{c}{\bf Full-history users} & \multicolumn{6}{c}{\bf Short-history users ($N=1$)} \\
\cmidrule(lr){2-7} \cmidrule(lr){8-13}
 & \multicolumn{3}{c}{\bf MIND} & \multicolumn{3}{c}{\bf EBNeRD} & \multicolumn{3}{c}{\bf MIND} & \multicolumn{3}{c}{\bf EBNeRD} \\
\cmidrule(lr){2-4} \cmidrule(lr){5-7} \cmidrule(lr){8-10} \cmidrule(lr){11-13}
\textbf{Method} & N@10 & N@5 & MRR & N@10 & N@5 & MRR & N@10 & N@5 & MRR & N@10 & N@5 & MRR \\
\midrule
NRMS  & \textbf{0.4256} & \textbf{0.3641} & \textbf{0.3322} & 0.5290 & 0.4629 & 0.4119 & 0.3536 & 0.2927 & 0.2748 & 0.4880 & 0.4127 & 0.3679 \\
NAML  & 0.4247 & 0.3633 & 0.3320 & \textbf{0.5312} & \textbf{0.4649} & \textbf{0.4134} & 0.3301 & 0.2908 & 0.2723 & 0.4889 & 0.4134 & 0.3690 \\
LSTUR & 0.3794 & 0.3183 & 0.2943 & 0.5114 & 0.4407 & 0.3909 & 0.2806 & 0.2205 & 0.2168 & 0.4496 & 0.3655 & 0.3305 \\
\midrule
\textbf{LSNR} & 0.4234 & 0.3620 & 0.3310 & 0.5200 & 0.4547 & 0.4027 & \textbf{0.3626}\rlap{$^\ast$} & \textbf{0.3015}\rlap{$^\ast$} & \textbf{0.2819}\rlap{$^\ast$} & \textbf{0.5022}\rlap{$^\ast$} & \textbf{0.4317}\rlap{$^\ast$} & \textbf{0.3824}\rlap{$^\ast$} \\
\bottomrule
\end{tabular}
\end{table*}

In the standard full-history setting, LSNR is a competitive alternative to dense baselines. On MIND, LSNR performs on par with the strongest dense models, sitting within half a percent of both NRMS and NAML on every metric while clearly outperforming LSTUR. On EBNeRD the picture is slightly different: LSNR trails the strongest dense models, NRMS and NAML by 2 to 3\% on NDCG@10 while remaining well ahead of LSTUR. These experiments show that LSNR captures semantic matching close to that of dense models.

Importantly, LSNR excels for users with short histories. As shown in the short-history columns of \autoref{tab:dense_comparison}, LSNR significantly outperforms all dense baselines when history is scarce ($N=1$). On MIND, LSNR surpasses the strongest dense baseline, NRMS, by 2.5\%. On EBNeRD, the gap is larger, with LSNR exceeding the strongest dense baseline, NAML, by 2.7\%. This trend is confirmed in \autoref{fig:history_length_performance}: LSNR leads the dense methods at small $N$, with all methods converging around $N=5$. We attribute this to the stability of sparse lexical signals. While dense vectors can suffer from semantic drift when interaction data is limited~\cite{thakur2021beir}, LSNR's explicit term matching enables precise retrieval even from a single interaction.

These findings establish LSNR as a reliable alternative to dense models: competitive in standard settings and significantly superior in short-history scenarios.

\begin{figure}[b]
    \centering
    \includegraphics[width=0.49\linewidth]{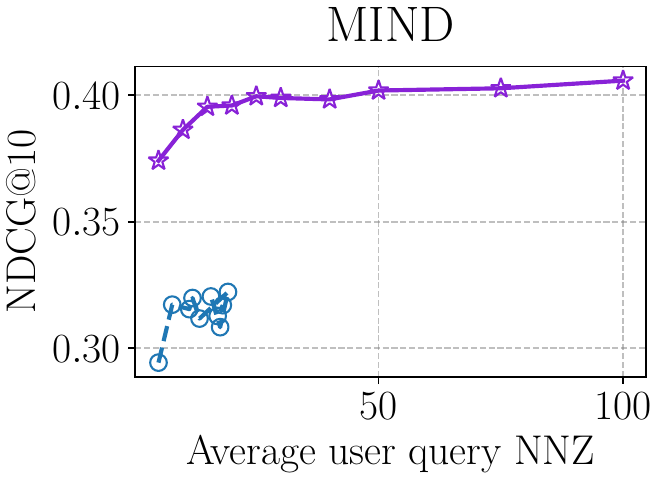}
    \includegraphics[width=0.49\linewidth]{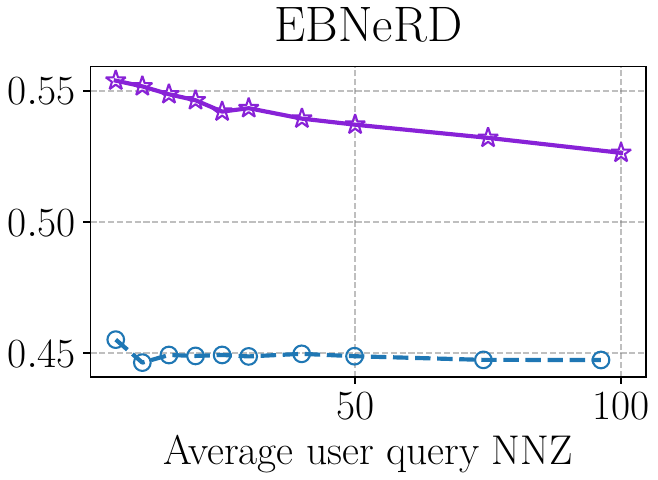}
    \includegraphics[width=0.5\linewidth]{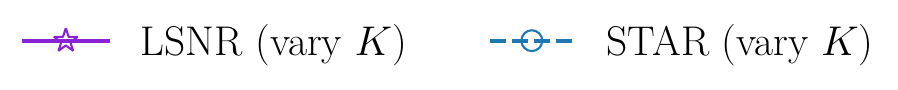}
    \caption{Accuracy-sparsity trade-off under user-vector pruning. Each curve sweeps the pruning threshold $K \in \{5, \dots, 100\}$; each point is one value of $K$, with smaller $K$ (sparser user vectors) toward the left. We plot NDCG@10 against average user-vector NNZ for MIND (left) and EBNeRD (right). NDCG@10 here is the impression-based metric, recomputed as the user vector is pruned to its top-$K$ terms.}
    \label{fig:tradeoffs}
\end{figure}

\subsection{Efficiency analysis}
\label{sec:efficiency}
 
We characterize LSNR's index footprint and retrieval cost, and analyze the effect of user-vector pruning on accuracy. The footprint and latency measurements use full-catalog retrieval, scoring each user vector against the entire article catalog through the index to reflect deployment-time cost. Accuracy, including the pruning curves in Figure~\ref{fig:tradeoffs}, remains the impression-based NDCG@10 used throughout the paper. Because the inner product $\mathbf{u}^\top \mathbf{s}_c$ is dominated by the user vector's non-zero entries, we sweep a pruning threshold $K \in \{5, \dots, 100\}$ that restricts each user vector to its top-$K$ highest-scoring terms. \autoref{fig:tradeoffs} plots NDCG@10 against the user-vector number of non-zeros (NNZ). All sparse measurements use a Seismic index~\cite{bruch2024efficient} on a single CPU thread. We index dense NRMS embeddings with HNSW~\cite{malkov2018efficient} on the same hardware.

\subsubsection{Pruning the user vector}

As \autoref{fig:tradeoffs} shows, LSNR exhibits a clear accuracy-sparsity trade-off on MIND. Tightening the pruning threshold reduces accuracy. However, even under the most aggressive pruning, LSNR substantially outperforms STAR's best result. STAR's accuracy is flat across thresholds, because its user vectors on MIND are already extremely sparse (\autoref{tab:indexing_efficiency}, 15.6 NNZ on average).

On EBNeRD, the pattern reverses: tighter pruning improves LSNR's accuracy. This divergence stems from dataset characteristics. EBNeRD features longer histories and substantially longer articles, so aggregating sparse representations across the entire history accumulates a broader vocabulary. Here, aggressive pruning may act as a denoising filter that retains only the user's core interests. STAR's user vectors on EBNeRD are denser (30.0 NNZ, \autoref{tab:indexing_efficiency}), yet its accuracy still does not benefit from pruning. This suggests that concatenation produces a flat, undifferentiated user vector in which all terms contribute roughly equally, so trimming any of them is as likely to hurt as to help.

\begin{table}[b]
\centering
\caption{Indexing and retrieval efficiency on the MIND and EBNeRD validation sets. NNZ\textsubscript{d}/NNZ\textsubscript{u} are average non-zero entries per document and user vector (for NRMS, the dense embedding dimension). User vectors are reported at $K=30$, our default pruning threshold. Size is on-disk index size in MB, and Lat. is mean end-to-end latency in ms (user encoding plus index retrieval), reported for characterization rather than as a comparative advantage.}
\label{tab:indexing_efficiency}
\setlength{\tabcolsep}{3pt}
\begin{tabular}{@{}l rrrr rrrr@{}}
\toprule
 & \multicolumn{4}{c}{\textbf{MIND}} & \multicolumn{4}{c}{\textbf{EBNeRD}} \\
\cmidrule(lr){2-5} \cmidrule(lr){6-9}
\textbf{Model} & \textbf{NNZ\textsubscript{d}} & \textbf{NNZ\textsubscript{u}} & \textbf{Size} & \textbf{Lat.} & \textbf{NNZ\textsubscript{d}} & \textbf{NNZ\textsubscript{u}} & \textbf{Size} & \textbf{Lat.} \\
\midrule
NRMS  & 768.0 & 768.0 & 208.06 & 1.46 & 768.0 & 768.0 & 31.84 &  1.16 \\
STAR  &  17.6 &  15.6 &   8.78 & 7.25 & 440.4 &  30.0 & 33.55 & 26.78 \\
LSNR  &  22.8 &  30.0 &  11.33 & 3.09 &  31.6 &  30.0 &  2.41 &  2.61 \\
\bottomrule
\end{tabular}
\end{table}

\subsubsection{Index footprint}

The two sparse methods, LSNR and STAR, reach very different document-sparsity operating points, and the contrast is dataset-dependent. On MIND, both produce highly sparse documents (LSNR 22.8 vs.\ STAR 17.6 NNZ, \autoref{tab:indexing_efficiency}), so their index sizes are comparable (11.33 vs.\ 8.78 MB). On EBNeRD, however, STAR's documents are roughly $14\times$ denser than LSNR's (440.4 vs.\ 31.6 NNZ), despite both methods using the same SPLADE document encoder and the same FLOPS regularization $\lambda_n$. We attribute this asymmetry to a difference in the training signal rather than the encoder or regularizer. STAR's document encoder is optimized against a single concatenated user vector covering the entire history, which itself spans a broad vocabulary, pushing the document encoder toward dense, broad representations. Instead, LSNR optimizes each article representation against per-article gradients during user-vector aggregation, encouraging more focused encodings.

LSNR's approach leads to a compact index that is an order of magnitude smaller than the uncompressed NRMS dense index on both datasets (11.33 vs.\ 208.06 MB on MIND, 2.41 vs.\ 31.84 MB on EBNeRD). Relative to STAR, LSNR's index is comparable on MIND but roughly $14\times$ smaller on EBNeRD (2.41 vs.\ 33.55 MB), where STAR's dense documents inflate its index. We note that the dense baseline is uncompressed. Quantization would reduce its footprint, trading index size against retrieval accuracy; a fully quantized comparison is left to future work.

\subsubsection{Retrieval cost}

\autoref{tab:indexing_efficiency} reports end-to-end latency (user encoding plus index retrieval) at our default pruning threshold $K=30$.  Decomposing this on MIND, LSNR's user-encoding step takes 0.97 ms, roughly 7× faster than STAR's 6.86 ms and on par with the dense NRMS encoder (0.89 ms), because LSNR aggregates precomputed article representations rather than encoding the raw concatenated history at retrieval time. This also makes LSNR substantially faster end-to-end than STAR ($2.3\times$ on MIND, $10\times$ on EBNeRD). Relative to the dense HNSW baseline, LSNR's inverted-index retrieval is slower in absolute terms, reflecting the state of dense ANN infrastructure. LSNR's end-to-end latency remains within a few milliseconds per query at this corpus scale.

\subsection{Pathway ablation}
\label{sec:pathway-ablation}

The hybrid user encoder constructs $\mathbf{u}$ from two pathways: a lexical pathway $\mathbf{u}_{\text{lex}}$ that aggregates per-article sparse vectors with attention weights, and a semantic pathway $\mathbf{u}_{\text{sem}}$ that projects the attention-weighted dense user representation into sparse space (Section~\ref{sec:user-encoder}). We test whether each pathway contributes to performance by training variants that activate one pathway in isolation and comparing them against the full hybrid.

\autoref{tab:pathway_ablation} reports NDCG@10 for the three configurations on MIND and EBNeRD, in full-history and short-history ($N=1$) settings. The full hybrid outperforms both single-pathway variants in all four cells. The two pathways perform almost identically when used in isolation, differing by at most 0.014 NDCG@10 across the four settings, indicating that neither pathway is strictly more informative than the other. Yet combining them yields substantial gains over either alone: 11.9\% over $\mathbf{u}_{\text{lex}}$-only and 11.3\% over $\mathbf{u}_{\text{sem}}$-only on MIND full-history, 9.8\% and 6.8\% on EBNeRD full-history. The pathways thus encode complementary signals rather than redundant information.

\begin{table}[h]
\centering
\caption{Pathway ablation. We train variants of LSNR with only the lexical pathway $\mathbf{u}_{\text{lex}}$, only the semantic pathway $\mathbf{u}_{\text{sem}}$, or both (hybrid). Performance is NDCG@10 for full-history and short-history ($N=1$) scenarios. Bold denotes the best configuration per column.}
\label{tab:pathway_ablation}
\begin{tabular}{l c c c c}
\toprule
 & \multicolumn{2}{c}{\textbf{MIND}} & \multicolumn{2}{c}{\textbf{EBNeRD}} \\
\cmidrule(lr){2-3} \cmidrule(lr){4-5}
\textbf{Configuration} & \textbf{Full} & \textbf{Short} & \textbf{Full} & \textbf{Short} \\
\midrule
$\mathbf{u}_{\text{lex}}$ only & 0.3784 & 0.3565 & 0.4734 & 0.4717 \\
$\mathbf{u}_{\text{sem}}$ only & 0.3805 & 0.3578 & 0.4868 & 0.4848 \\
Both (hybrid)                  & \textbf{0.4234} & \textbf{0.3626} & \textbf{0.5200} & \textbf{0.5022} \\
\bottomrule
\end{tabular}
\end{table}

The hybrid's advantage is most pronounced in full-history settings, where there is more interaction data for the two pathways to capture distinct patterns from. At $N=1$, where the user is represented by a single article, the gap narrows on MIND (1.7\% and 1.3\% over the single-pathway variants) but remains substantial on EBNeRD (6.5\% and 3.6\%). This is consistent with the dataset characteristics: EBNeRD's longer articles provide more text per interaction for both pathways to draw on, so even a single article supports meaningful per-article aggregation and dense-projected expansion.

This result shows the advantage of LSNR's two-pathway design. $\mathbf{u}_{\text{lex}}$ provides direct lexical signal from individual articles, while $\mathbf{u}_{\text{sem}}$ supplies user-level semantic expansion. The two are roughly equivalent in standalone capacity but encode different aspects of the user's interests, and combining them captures both.

\subsection{User aggregation methods}
\label{sec:ablation}
Having established that both user encoding pathways contribute, we next ablate the aggregation function within the lexical pathway $\mathbf{u}_{\text{lex}}$, keeping the semantic pathway $\mathbf{u}_{\text{sem}}$ unchanged. We compare our additive attention against four alternatives: attention with positional encodings, mean pooling, temporal decay, and max pooling. We report NDCG@10 in full- and short-history ($N=1$) settings.

The results in \autoref{tab:ablation_results} show that our additive attention is the strongest aggregation across both benchmarks, justifying its use in the final architecture. Adding positional encodings degrades performance on both datasets. That order information actively hurts supports our decision to omit it, and matches prior findings that strict temporal order is a weak signal for news recommendation~\cite{wu2022news}. The parameter-free pooling strategies fall further behind, and do so unevenly across datasets: max pooling is the strongest alternative on MIND but the weakest on EBNeRD, while mean pooling and temporal decay are weakest on MIND and slightly stronger on EBNeRD. No single parameter-free strategy is robust across both the short title-and-abstract articles of MIND and the long-form articles of EBNeRD, whereas learned attention is consistently strongest in every setting, which is what motivates its use in LSNR.

\begin{table}[h]
\centering
\caption{Aggregation-function ablation on MIND and EBNeRD, holding the semantic pathway $\mathbf{u}_{\text{sem}}$ constant. Performance is NDCG@10 for both full-history and short-history ($N=1$) scenarios. Bold denotes the best method per column.}
\label{tab:ablation_results}
\begin{tabular}{l c c c c}
\toprule
 & \multicolumn{2}{c}{\textbf{MIND}} & \multicolumn{2}{c}{\textbf{EBNeRD}} \\
\cmidrule(lr){2-3} \cmidrule(lr){4-5}
\textbf{Method} & \textbf{Full} & \textbf{Short} & \textbf{Full} & \textbf{Short} \\
\midrule
Attention         & \textbf{0.4234} & \textbf{0.3626} & \textbf{0.5200} & \textbf{0.5022} \\
Attention + PE    &        0.4086   &        0.3582   &        0.5120   &        0.4917   \\
Average pooling   &        0.4020   &        0.3513   &        0.5082   &        0.4895   \\
Temporal decay    &        0.3990   &        0.3516   &        0.5048   &        0.4920   \\
Max pooling       &        0.4138   &        0.3575   &        0.4996   &        0.4843   \\
\bottomrule
\end{tabular}
\end{table}

\section{Conclusion}
We have introduced learned sparse news recommendations (LSNR), a framework that adapts learned sparse retrieval to the long-form text and user histories of the news domain. By encoding each article independently and aggregating them via a learned, attention-based user encoder, LSNR eliminates the context-window bottleneck and the interest collapse that concatenation induces in prior learned sparse methods such as STAR.

Our evaluation on the MIND and EBNeRD benchmarks yields three findings. First, LSNR significantly outperforms both BM25 and STAR by representing users at the granularity of individual articles. Second, this advantage persists in short-history settings ($N=1$) where STAR processes the same input without truncation, showing that LSNR's gains stem from improved per-article learning rather than from avoiding truncation. Third, LSNR performs competitively with state-of-the-art dense recommenders (NRMS, NAML) in standard settings and significantly outperforms them when user history is short, while serving from an index an order of magnitude smaller than the uncompressed dense baseline on both datasets.

We identify several limitations of our work. First, LSNR's advantage over dense models is concentrated in the short-history setting: in the full-history setting, it matches the strongest dense recommenders on MIND but trails them by 2 to 3\% on EBNeRD. Second, on the recall-latency frontier, a well-tuned dense HNSW index matches or exceeds LSNR's retrieval accuracy at lower query latency, so LSNR's efficiency benefit is its compact index rather than retrieval speed. Third, our evidence comes from two benchmarks in two languages, and the datasets are the ``small'' versions. Generalization to larger corpora and other domains remains for future work. Fourth, on MIND, the two sparse models start from different checkpoints, an asymmetry that favors STAR and that we therefore read as making LSNR's gain conservative, though it leaves the comparison short of fully controlled.

Beyond textual signals, auxiliary features such as article categories or user demographics could further sharpen the sparse user vector. Because each dimension of the user vector maps to a vocabulary term, the representation could in principle support term-level inspection or editing of user interests. We leave a direct evaluation of such controls to future work.

\section*{Resources}
At \url{https://github.com/davidvos/learned-sparse-news-recs} we share the following resources, under an MIT license, to facilitate reproducibility:
\begin{itemize}[leftmargin=*]
  \item Our proposed model, LSNR, with its attentive user and news encoders, and all baselines: STAR and the dense recommenders NRMS, NAML, and LSTUR.
  \item Preprocessing and pre-tokenization pipelines for the MIND and EBNeRD datasets.
  \item Training and evaluation scripts reporting performance metrics and sparsity, plus an efficiency benchmark for index size and retrieval latency.
  \item A central configuration file with all hyperparameters, fixed seeds, and chronological splits, and documentation to reproduce the reported results.
\end{itemize}

\begin{acks}
    This research was (partially) supported by the Dutch Research Council (NWO), under project numbers 024.004.022, NWA.1389.20.\-183, KICH3.LTP.20.006, and VI.Vidi.223.166, and the European Union under grant agreement No. 101201510 (UNITE). Views and opinions expressed are those of the author(s) only and do not necessarily reflect those of their respective employers, funders and/or granting authorities.
\end{acks}

\bibliographystyle{ACM-Reference-Format}
\balance
\bibliography{references.bib}

\end{document}